# Analysis of a distributed fiber-optic temperature sensor using single-photon detectors

**Shellee D. Dyer[1,*], Michael G. Tanner[2], Burm Baek[1], Robert H. Hadfield[2], and Sae Woo Nam[1]**

[1]*National Institute of Standards and Technology, Optoelectronics Division, Boulder, CO 80305 USA*

[2]*School of Engineering and Physical Sciences, Heriot-Watt University, Edinburgh EH14 4AS U.K.*

[*]*sdyer@boulder.nist.gov*

**Abstract:** We demonstrate a high-accuracy distributed fiber-optic temperature sensor using superconducting nanowire single-photon detectors and single-photon counting techniques. Our demonstration uses inexpensive single-mode fiber at standard telecommunications wavelengths as the sensing fiber, which enables extremely low-loss experiments and compatibility with existing fiber networks. We show that the uncertainty of the temperature measurement decreases with longer integration periods, but is ultimately limited by the calibration uncertainty. Temperature uncertainty on the order of 3 K is possible with spatial resolution of the order of 1 cm and integration period as small as 60 seconds. Also, we show that the measurement is subject to systematic uncertainties, such as polarization fading, which can be reduced with a polarization diversity receiver.

**OCIS codes:** (040.3780) Low light level detectors; (060.2370) Fiber optics sensors; (120.4825) Optical time domain reflectometry; (190.5650) Raman effect

---

#### References and links

---

## 1. Introduction

Distributed fiber-optic sensors are an attractive alternative to multiplexed point sensors, because a single fiber-optic cable can potentially replace thousands of individual sensors, dramatically simplifying sensor installation and readout. Two important classes of distributed sensors are optical frequency domain reflectometry (OFDR) [1] and optical time-domain reflectometry (OTDR) [2]. OFDR systems have the advantage of high spatial resolution (on the order of 1 mm) [1], but the maximum range is generally limited to tens of meters. Additionally, OFDRs are generally based on interferometric techniques, and special care is required to avoid measurement problems with vibration and polarization. OFDR systems have been implemented with both single-mode and multi-mode fiber under test (FUT).

OTDR measurements typically employ a pulsed source and determine position information from the time-of-flight of the photons that are backscattered from the fiber under test. One important class of OTDR sensors is based on measurements of Raman scattering. Dakin et al. were the first to demonstrate that the temperature could be determined from a ratio of measured Stokes and anti-Stokes Raman scattering [2]. The range of the OTDR sensors is typically limited to about 10 km by losses in the fiber and fiber intermodal dispersion (if multi-mode fiber is used as the sensing fiber). Another factor affecting the OTDR range is the repetition rate of the pump laser, but this can be modified with modulation, either internal or external to the pump laser. The spatial resolution of the OTDR is determined by the convolution of the laser's pulse width with the response function of the detection system. Typical performance of a Raman-based OTDR is 1 m spatial resolution, and 1 K temperature uncertainty for a five-minute measurement integration period, with a sensing range of 10 km [3].

Raman-based fiber-optic distributed temperature sensors have found a strong commercial niche in energy industries, where they provide thermal profiling of gas and oil wells [4]. Fiber sensors, with their small size, low weight, and immunity to electromagnetic interference, also have applications in monitoring nuclear reactors [5]. One drawback of distributed fiber sensors is that the backscattered signal is typically very weak; therefore, early work employed multi-mode fibers to increase the collection of backscattered photons. Additionally, pump lasers with wavelengths near 800 nm - 900 nm were chosen so that high-performance silicon avalanche photodiodes (APDs) could be used for detection [2].

The range of the OTDR-based sensors is limited by losses in the fiber and intermodal dispersion; hence, one key improvement for these sensors is the replacement of multimode fiber with inexpensive telecom-grade low-loss single-mode fiber. Bolognini et al. have demonstrated a measurement distance range greater than 40 km in single-mode dispersion-shifted fiber, using optical amplification in combination with coded pulses [6]. In that example, the temperature resolution was 5 K, and the spatial resolution was limited to 17 m by the timing resolution of the InGaAs APDs used for detection.

In a previous paper, we demonstrated a distributed temperature sensor with spatial resolution on the order of 1 cm in standard telecom single-mode fiber [7]. We are able to achieve this high-spatial resolution in a single-mode FUT at wavelengths near 1550 nm as a result of the unique properties of our superconducting nanowire single-photon detectors (SNSPDs) [8]. We employed time-correlated single-photon counting techniques [9-10], in which we created histograms of the time delays between the launch of a laser pulse and the detection of backscattered photons at the SNSPDs. SNSPDs have exceptional performance in the visible to mid-infrared wavelength regions, with single-photon sensitivity, low dark counts, short recovery periods, and low timing jitter [8]. The timing jitter of the SNSPDs is on the order of 65 ps, enabling distributed temperature sensing with centimeter spatial resolution at wavelengths near 1550 nm. In this paper, we provide a detailed analysis of the measurement uncertainties. The measurement uncertainty is affected by both statistical uncertainties, arising from the counting of single photons, and systematic uncertainties, resulting from factors such as bends and twists in the fiber.

## 2. Theoretical Model of Raman Scattering in Fiber

We model the detector count rates from Raman backscattering as follows [11]:

$$I_u = \eta_u \Delta \nu_u P_0 L \left| g_{R,u} \right| \mathcal{N}\left(\Omega_{up}\right) D_c + B_u, \tag{1}$$

where $I_u$ is the number of backscattered photons per second, with the subscript $u$ replaced by either $s$ or $a$ for Stokes or anti-Stokes scattering, $B_u$ is the background photon rate resulting from dark counts and Rayleigh-backscattered amplified spontaneous emission (ASE), $\eta_u$ is a product of the detection efficiency (DE) of the SNSPDs and the transmission coefficient of the filter chains, $\Delta\nu_u$ is the bandwidth of the filters applied to the Stokes and anti-Stokes channels, $P_0$ is the peak pump power, $L$ is fiber length, $g_{R,u}$ is the Raman gain coefficient, $D_c$ is the duty cycle of the pump signal, $\Omega_{up} = \Omega_u - \Omega_p$ is the radial frequency detuning between the pump and the Stokes or anti-Stokes wavelengths, and $\mathcal{N}$ represents the phonon population, as follows:

$$\mathcal{N} = \begin{cases} \dfrac{1}{\exp\left(\dfrac{\hbar\left|\Omega_{up}\right|}{k_B T}\right) - 1} & when \quad \Omega_{up} > 0 \\ \dfrac{1}{\exp\left(\dfrac{\hbar\left|\Omega_{up}\right|}{k_B T}\right) - 1} + 1 & when \quad \Omega_{up} < 0. \end{cases} \tag{2}$$

If we approximate the Raman gain coefficient $g_{R,u}$ and detuning $\Omega_{up}$ to be independent of frequency over our Stokes and anti-Stokes filter bandwidths, and assume $\Omega_{ap} \approx \Omega_{sp}$, then the ratio of Stokes to anti-Stokes photon count rates is given by

$$\frac{I_s(x) - B_s}{I_a(x) - B_a} = C \exp\left(\frac{\hbar\left|\Omega_{sp}\right|}{k_B T(x)}\right), \tag{3}$$

where $I_s$ is the Stokes count rate, $B_s$ is the Stokes background count rate, $I_a$ is the anti-Stokes count rate, $B_a$ is the anti-Stokes background count rate, $x$ is the position in the FUT, and $C$ is a constant. This equation provides a model from which we can calculate the temperature of the FUT from measurements of the Raman scattering. In this paper we determine the constant $C$ from a reference measurement at a known temperature, but it is also possible to eliminate the need for a reference measurement through careful characterization of filter bandwidths, filter losses, Raman gain coefficient, and detection efficiency. In that case, this temperature sensing system could potentially be employed as a primary thermometer with the ability to accurately determine temperatures from first principles.

## 3. System Description

A diagram of our measurement is shown in Fig. 1. Our light source is a femtosecond fiber laser with a repetition frequency of 36 MHz and a spectrum centered near 1550 nm. Our entire measurement system was constructed from inexpensive, off-the-shelf fiber optic components that were designed for the telecommunication industry. The sensing fiber is a standard low-loss single-mode telecommunications fiber with a mode field diameter of 10.4 μm and attenuation less than 0.2 dB/km at a wavelength of 1550 nm. We filter the pump laser to a 1 nm linewidth at a wavelength of 1533.47 nm (ITU Ch.55). We attenuate the pump laser before amplifying it with an erbium doped fiber amplifier (EDFA). The attenuator is used to minimize pulse distortion due to gain saturation in the EDFA. The pump signal is again filtered post-amplification to remove the broadband ASE from the pump, which could otherwise be Rayleigh backscattered in the FUT and create unwanted counts at our detectors. We apply multiple filters in series both before and after the EDFA, because each filter provides only 20-30 dB rejection of unwanted wavelengths. We inject the pump light into our FUT via a fiber-optic circulator. The length of our fiber under test is currently limited to about 2.8 m by the repetition rate of our pump laser. We expect that this technique could easily be extended to test fibers with lengths on the order of kilometers with suitable choice of pump laser or with a pulse-picking modulator applied to our pump laser. The light that is backscattered from the FUT is directed by the circulator to a series of filters that have been chosen to reject the pump wavelength. Without these filters, we would have unwanted photons at the pump wavelength reaching our detector due to Rayleigh backscattering in the FUT, Fresnel reflection at the endface of the FUT, and imperfections in the circulator and fiber-to-fiber interconnections. The 980 nm filters are used to block any of the EDFA pump photons that might otherwise reach the detectors. We use a series of band splitters to separate and filter the Raman backscattered photons into an S-band channel (1460 nm – 1490 nm) and an L-band channel (1570 nm – 1610 nm). The pump wavelength was chosen to be approximately equally spaced between the S- and L-bands. The light from the L- and S-band channels are coupled to SNSPDs. The time-interval analyzer (TIA) creates histograms of the time delays between a laser clock electrical pulse and the electrical pulses created by the detection of photons at the SNSPDs.

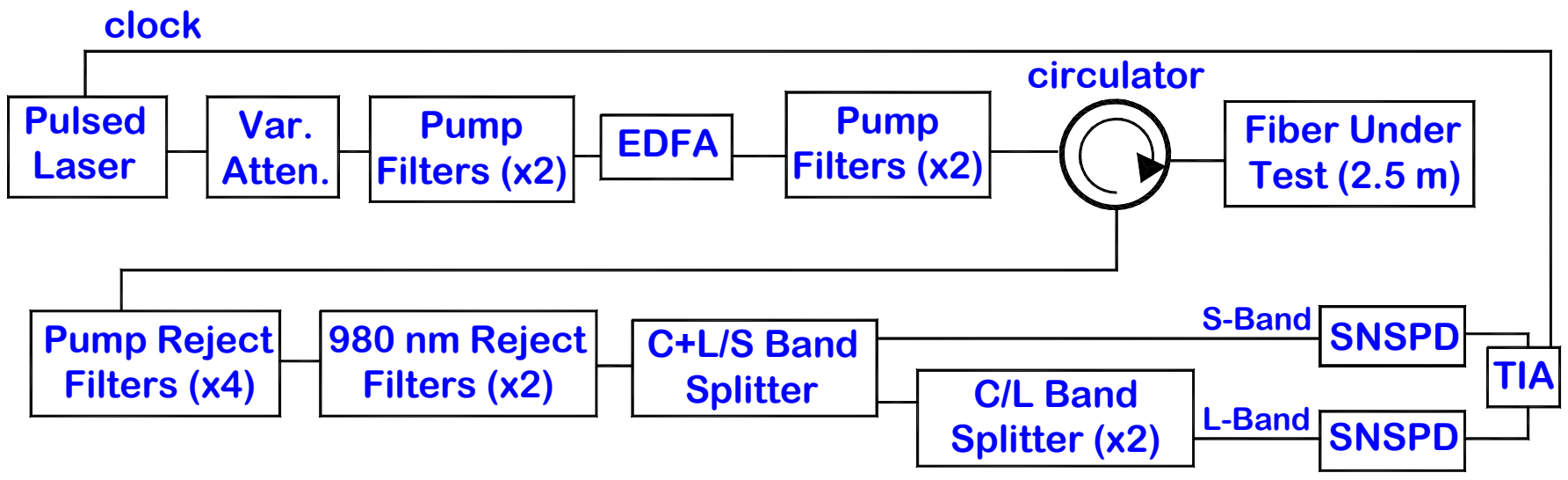

Fig. 1. Diagram of our high-spatial-resolution fiber-optic temperature sensor system. EDFA: erbium-doped fiber amplifier, SNSPD: superconducting nanowire single-photon detector, TIA: time interval analyzer. The pulsed laser is a femtosecond fiber laser with a 36 MHz clock rate. The pump filters are 1 nm linewidth at a wavelength of 1533.47 nm. The pump reject filters are identical to the pump filters, but are connected so that they filter out the pump wavelength. The C+L/S bandsplitter separates the S-band wavelengths (1460 nm -1490 nm) from the C-band wavelengths (1530 nm -1565 nm) and L-band wavelengths (1570 nm -1610 nm). The C/L bandsplitter separates the L-band wavelengths from the C-band wavelengths.

The Raman gain coefficient ($g_{R,u}$) in silica fibers generally increases with detuning, reaching a peak near 15 THz (approximately 100 nm), but is typically quite small, i.e., on the order of 1.0 $W^{-1}\cdot km^{-1}$. Although SNSPDs with high system detection efficiency (DE) have been demonstrated elsewhere [12-13], the ones that we used in this project had relatively low DEs of 0.6 % and 1 %. In addition to the DE, we must also account for the insertion losses of all filters between the FUT and the detectors, as well as the insertion loss of the circulator; this yields a filter chain insertion loss on the order of 10 dB for both the S- and L-band channels. Our pump power was limited by gain saturation in our EDFA to an average power of approximately 18 mW. Therefore, our options for increasing count rates are to increase the bandwidth of our Stokes and anti-Stokes filters or to increase the detuning of our Stokes and anti-Stokes filters. We wanted to use inexpensive standard telecom components, which led to our choice of 1533.47 nm for the pump wavelength, Stokes filter range of 1570 nm - 1610 nm, and anti-Stokes filter range of 1460 nm - 1490 nm.

## 4. Experimental Results

Distributed temperature sensors are well-suited for measuring one-dimensional structures, such electrical wires, two-dimensional structures, such as a parking lot or roof, and three-dimensional structures, such as a building. We demonstrated a two-dimensional temperature sensor by taping our sensing fiber to a cardboard template in a regular meander pattern, as shown in Fig. 2. This layout is an example of how a distributed fiber temperature sensor might be applied to a two-dimensional structure, such as the surface of a bridge.

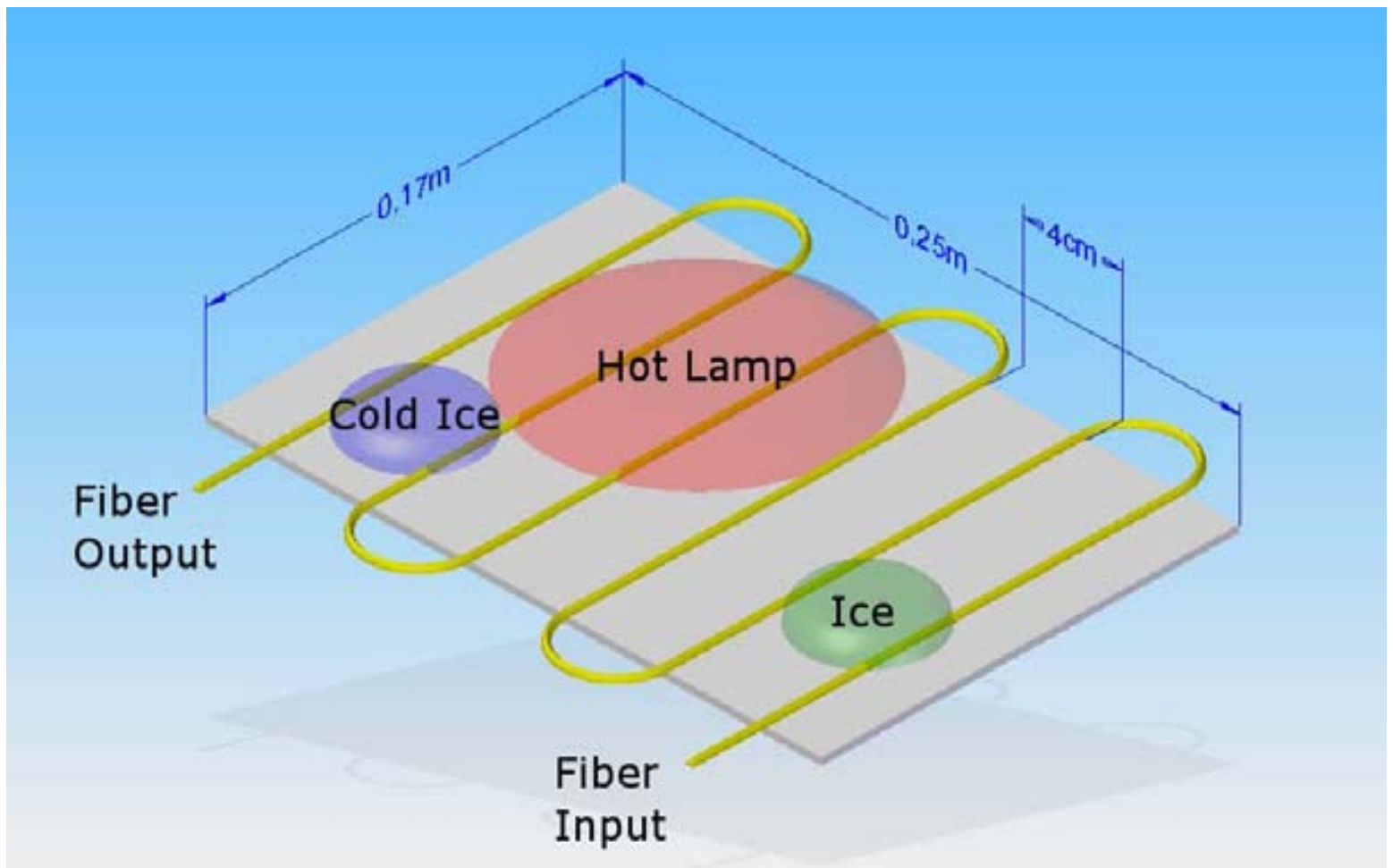

Fig. 2. Diagram of our layout for demonstration of a two-dimensional distributed fiber temperature sensor. Our sensing fiber is taped in a regular meander pattern to a cardboard box with 4 cm spacing between each horizontal run of the meander.

We heated and cooled separate sections of our two-dimensional sensing fiber meander using a heat lamp and two ice trays. The two ice trays were briefly immersed in liquid nitrogen to cool them to temperatures below 273 K. We recorded histograms of time delays between the laser clock pulses and the output pulses of our SNSPDs for both the Stokes and anti-Stokes channels, as described above. Integration period was 5 minutes for this example. We modeled the histogram counts per bin due to Raman scattering as

$$\mathrm{N}_u[m] = I_u[m]t_{\mathrm{int}} \approx \eta_u \Delta \nu_u P_0 L_{bin} \left| g_{R,u} \right| \mathcal{N}[m] D_c t_{\mathrm{int}}, \tag{4}$$

where $t_{\text{int}}$ is the histogram integration period, and $L_{\text{bin}} = ct_{bin}/2n_{fiber}$ is the equivalent length of the histogram time bins ($t_{bin}$), rather than the total fiber length. As described above, we have approximated the phonon distribution $\mathcal{N}$ and the Raman gain $g_{R,u}$ as independent of wavelength over the linewidth of our filters. The bin variable $m = x/L_{bin}$ represents the position in the fiber. By comparison, Eq. (1) gives Raman scattering in terms of the number of photons per second, while Eq. (4) describes the total number of photon counts. Although we did perform some measurements with different histogram bin sizes, all data shown here were measured with $t_{bin} =$ 64 ps, chosen because our SNSPD timing jitter is on the order of 65 ps FWHM. Histogram counts resulting from background effects such as dark counts and ASE can be modeled as $\mathrm{N}_{Bu} = B_u t_{int}$. The temperature of the fiber can be calculated from the following modification of Eq. (3):

$$T[m] = \frac{\hbar\left|\Omega_{sp}\right|}{k_B\left[\ln\left(\mathrm{N}_s[m]-\mathrm{N}_{Bs}\right)-\ln\left[C\left(\mathrm{N}_{as}[m]-\mathrm{N}_{Bas}\right)\right]\right]}. \tag{5}$$

We calculated the temperature as a function of position in the fiber, and then reconstructed the temperature of our two-dimensional surface. The result is shown in Fig. 3, where the locations of the heat lamp and the two ice trays are clear from the measured temperatures.

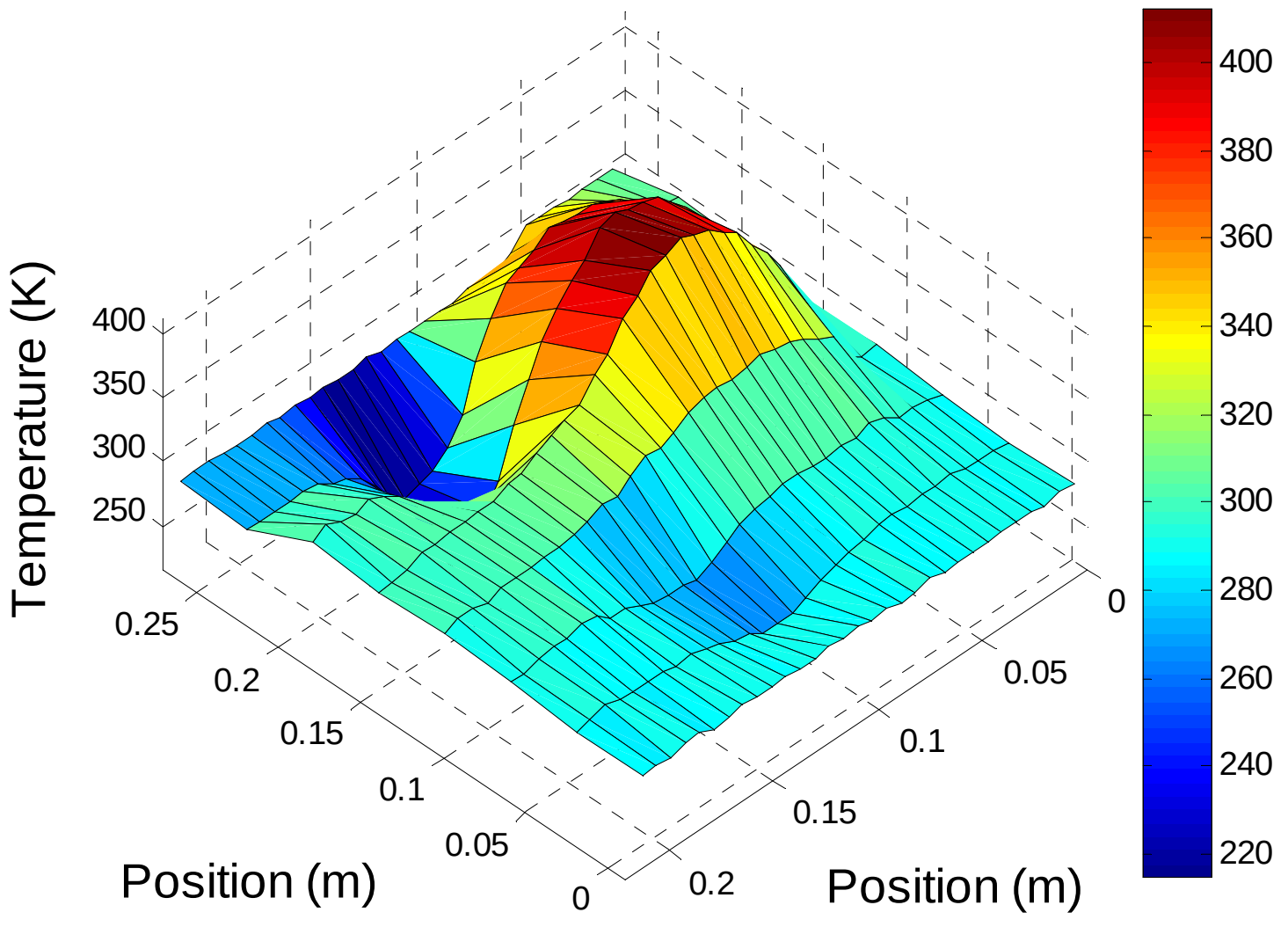

Fig. 3. Temperature of a two-dimensional fiber meander. The crossings of the mesh are actual data points, while the color scale represents the interpolation of temperatures between actual data points. For this measurement, one section of the two-dimensional meander was illuminated with a heat lamp, and two sections of the meander were covered with ice trays that had been immersed in liquid nitrogen to cool them below the freezing point of water.

## 5. Uncertainty Analysis

By use of standard uncertainty analysis techniques [14], the uncertainty of the temperature measurement can be expressed as

$$\frac{\Delta T_{meas}}{T_{meas}} \approx \frac{kT_{meas}}{\hbar\Omega_{sp}} \sqrt{\left(\frac{1}{\mathrm{N}_s}\right)+\left(\frac{1}{\mathrm{N}_{as}}\right)+\left(\frac{\Delta C}{C}\right)^2}, \tag{6}$$

where $\Delta C$ is the uncertainty of the reference constant $C$, and we have assumed that the one-sigma uncertainties $\Delta \mathrm{N}_s$ and $\Delta \mathrm{N}_{as}$ of the measured histogram counts in the Stokes and anti-Stokes channels are those of counting experiments, i.e., one-sigma uncertainty of $\Delta \mathrm{N} = \mathrm{N}^{1/2}$. We have also assumed that the background counts in the Stokes and anti-Stokes channels are negligible. This approximation is reasonable for all temperatures except cryogenic temperatures, for which the background counts would need to be included in the analysis.

Eq. (6) provides some insight into the characteristics of the temperature uncertainty. The Stokes and anti-Stokes histogram counts $\mathrm{N}_s$ and $\mathrm{N}_{as}$ increase linearly with integration period. For short integration periods, the first two terms of Eq. (6) will dominate, and increasing the integration period will reduce the uncertainty of the measurement. However, for long integration periods, the third term of Eq. (6) will dominate, and further increases in the integration period will yield no improvement in the uncertainty of the temperature measurement. For all integration periods, the uncertainty of the measured temperature is always limited by the calibration uncertainty $\Delta C$. This measurement system is a secondary thermometer if $C$ is determined from a reference measurement at a known temperature, and that is the approach that we used in this paper. Alternatively, this measurement system could potentially be a primary thermometer, if the constant $C$ is determined with low uncertainty through a careful characterization of all system parameters.

The analysis leading to Eq. (6) has neglected the effects of losses in the sensing fiber. This is reasonable for the short lengths of sensing fiber used in this paper, but for longer fibers the losses can have a significant effect on the uncertainty of the measurement. To account for losses in the fiber, we multiply the $\mathrm{N}_s$ and $\mathrm{N}_{as}$ terms by a position-dependent factor of $\exp(-2\alpha x)$, where $\alpha$ is the fiber loss and $x$ is the position along the fiber. Our single-mode fiber has attenuation less than 0.2 dB/km at wavelengths near 1550 nm, so the loss is negligible for sensing fiber lengths shorter than 1 km.

Another notable point about the analysis leading to Eq. (6) is that this analysis has focused on random uncertainties. This measurement is also affected by systematic uncertainties, such as the effects of polarization discussed in Sec. 6 of this paper. Other systematic uncertainties include fluctuations in the laser power and/or ASE output. If this system were to be used as a primary thermometer, these systematic uncertainties must be well characterized.

To demonstrate the low uncertainty of our measurement as a secondary thermometer, we recount an experiment first described in Ref. [7]. We immersed a section of the sensing fiber and a comparison thermocouple in close proximity in a hot water bath. We then recorded repeated histograms, each integrated for 60 seconds, as the water bath cooled. For each histogram, we also recorded the corresponding reading of the thermocouple. In Fig. 5 we show a comparison of the temperatures measured by the thermocouple and the temperatures determined by a section of the sensing fiber that was completely immersed in the water bath. We also show the estimated uncertainty obtained from Eq. (6). This integration period is sufficiently short that the third term of Eq. (6) can be neglected. The temperature uncertainty is on the order of 3 K for a 60-second integration period. In [7] we demonstrated that it is possible to trade spatial resolution for better temperature uncertainty by performing spatial averaging of the data, obtaining 1 K temperature uncertainty at a spatial resolution of 5 cm.

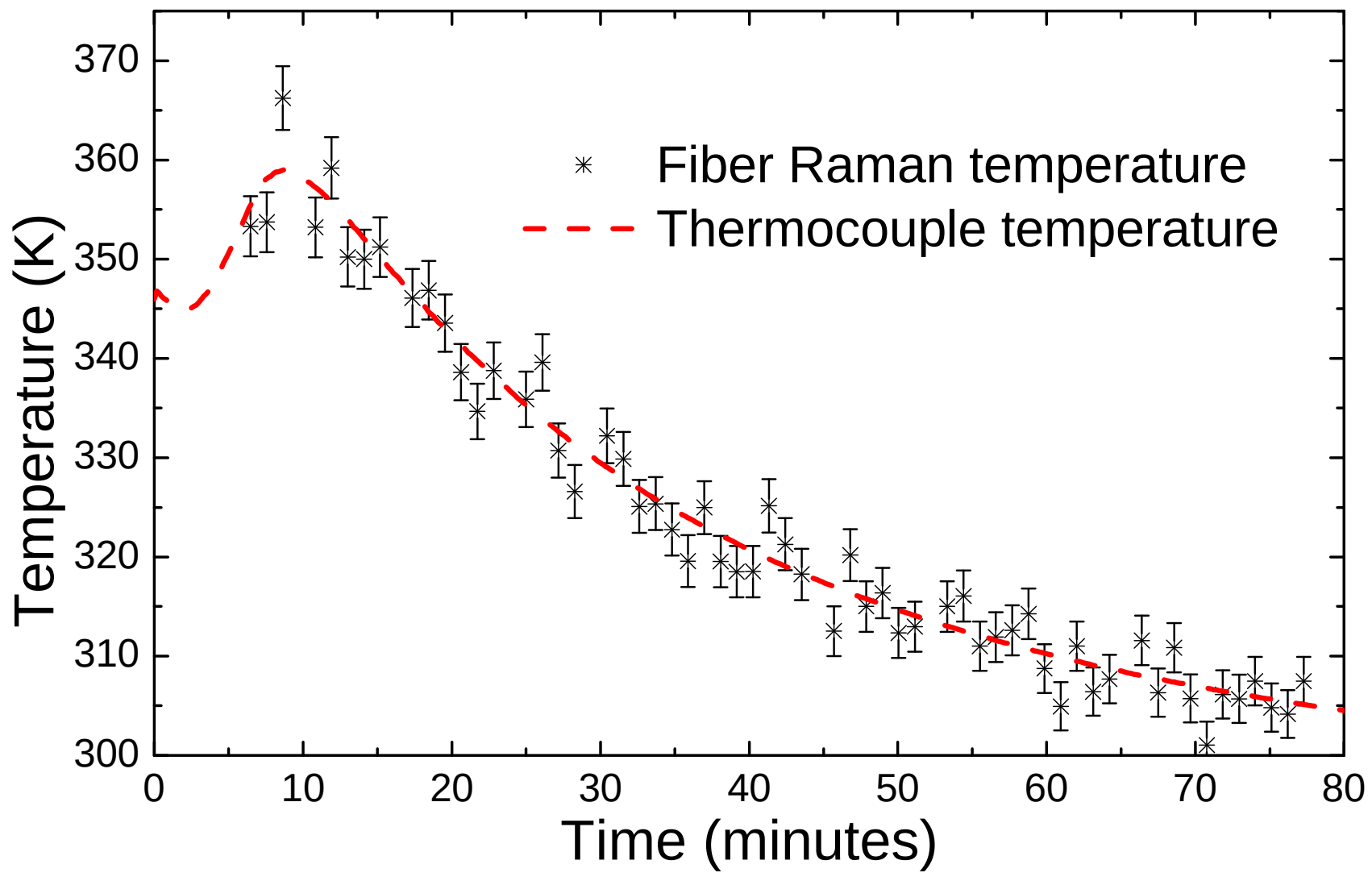

Fig.4. Measured temperature of the fiber Raman sensor compared with the corresponding temperature determined with a thermocouple. Data are from an experiment in which both the fiber and the thermocouple were immersed in a hot water bath, and measurements were taken every 60 seconds as the water bath slowly cooled over a 1 hour period. Each fiber sensor measurement shown is determined from a 1-minute integration period. The error bars show the uncertainty calculated from Eq. (6).

## 6. Effects of Polarization

For the highest-accuracy temperature measurements, it is important to account for the effects of polarization in distributed fiber-optic Raman sensors. The state of polarization of the pump laser varies randomly with position in the sensing fiber, as a result of localized variations in the birefringence of the core created by bends or twists in the fiber. Additionally, the Raman gain, which is often approximated as a scalar, is more accurately modeled as a tensor, with the Raman scattering that is co-polarized with the pump distinctly stronger than the cross-polarized Raman scattering [15-16]. The combination of pump polarization that varies with position in the sensing fiber and polarization-dependent Raman gain yields a measured Raman scattering strength that varies with position in the fiber. This effect will not be obvious in the case of sensors with spatial resolution on the order of 1 m, because the localized polarization-induced variations in Raman scattering are effectively washed out by the measurement. With higher spatial resolution, the polarization effects become apparent. The SNSPDs also contribute polarization sensitivity as a result of the nanowire geometry [17]. At wavelengths in the telecom C-band, the detection efficiency of these devices can be twice as large for polarizations that are parallel to the meander wires compared with polarizations that are perpendicular to the meander [17].

One solution to the problem of polarization sensitivity is a polarization diversity receiver (PDR). These are often applied to fiber-optic interferometers to avoid problems with polarization fading [5]. A diagram of the Raman sensor system with PDRs included in both the S-band channel and the L-band channel is shown in Fig. 5. We use fiber-optic polarizing beam splitters (PBS) to separate the S-band and L-band channels into "H" and "V" polarization components. We also use fiber-optic polarization controllers to align the output states of the PBS with the favored polarizations of the SNSPDs. For comparison, it may be helpful to note that polarization controllers are not used in Fig. 1, because in that case the state of polarization at the SNSPDs is not constant, rather it varies with the position of backscattering of photons in the FUT. We did not have four SNSPDs available for the PDR experiment, so we demonstrated a proof-of-principle example by measuring the S-band histograms separately from the L-band components.

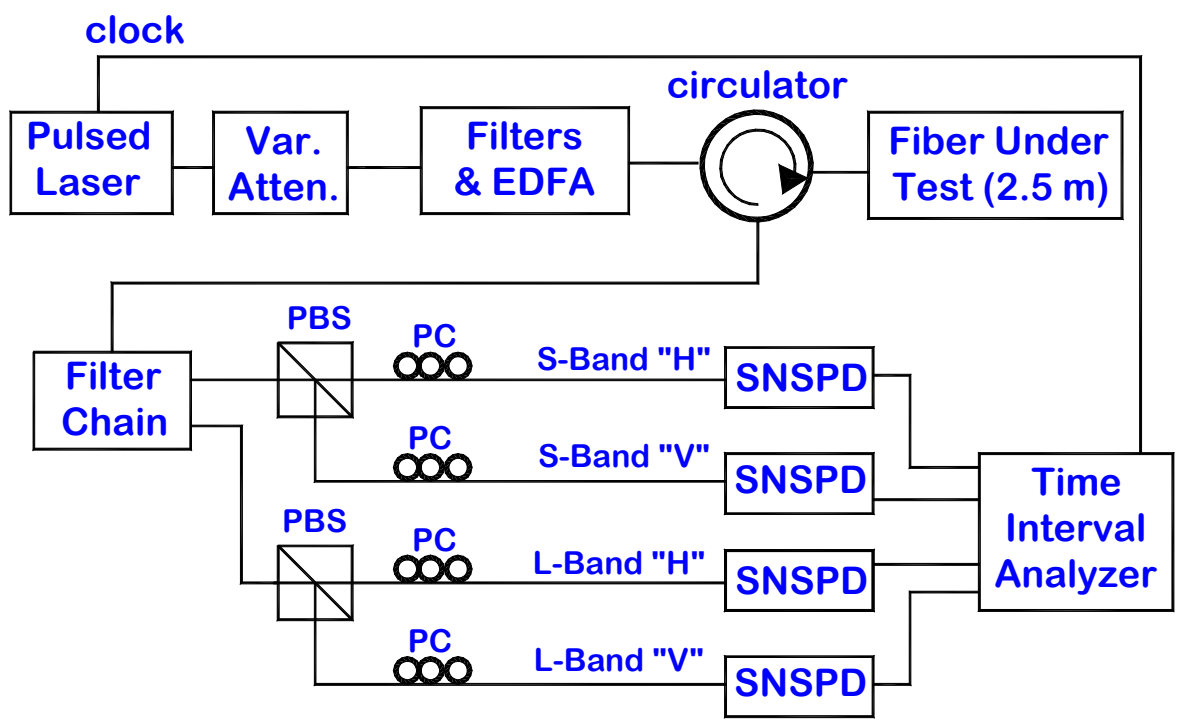

Fig. 5. Diagram of the Raman sensor measurement system with polarization diversity receivers (PDRs) included on the S-band and L-band receiver channels. EDFA: erbium doped fiber amplifier; PC: polarization controller; PBS: polarizing beam splitter; SNSPD: superconducting nanowire single-photon detector.

The two-dimensional fiber meander setup of Fig. 2, with its periodically spaced fiber bends, provided an interesting test case for our PDR. The sharp bends of this fiber meander create an extreme example of the effects of bend-induced changes in the birefringence of the fiber core. This sensing fiber, when held at room temperature, should not yield large variations in histogram counts versus position in the fiber, and the temperature calculated from the histograms should be fairly consistent throughout the length of the fiber under test. In Fig. 6 we show the raw histogram data sets for the sensing fiber held at room temperature; in Fig. 6(a) are the histograms for the two polarizations of the S-band channel, and in Fig. 6(b) are the histograms for the two polarizations of the L-band channel. The integration period for these histograms was 5 minutes. Near a fiber position of 2.5 m the histogram counts for both polarizations drop dramatically; this occurs when the delay between laser clock pulses and detector clicks is longer than the round trip period in the fiber under test.

For delays beyond this point, the histogram counts result from both the ASE that occurs between laser pulses as well as detector dark counts. At a delay near 2.8 m (corresponding to the laser clock rate of 36 MHz), the histogram counts drops to zero. This occurs because the subsequent laser clock pulse has reset the time-interval analyzer measurement, so that the maximum measurable delay is limited by the laser clock period.

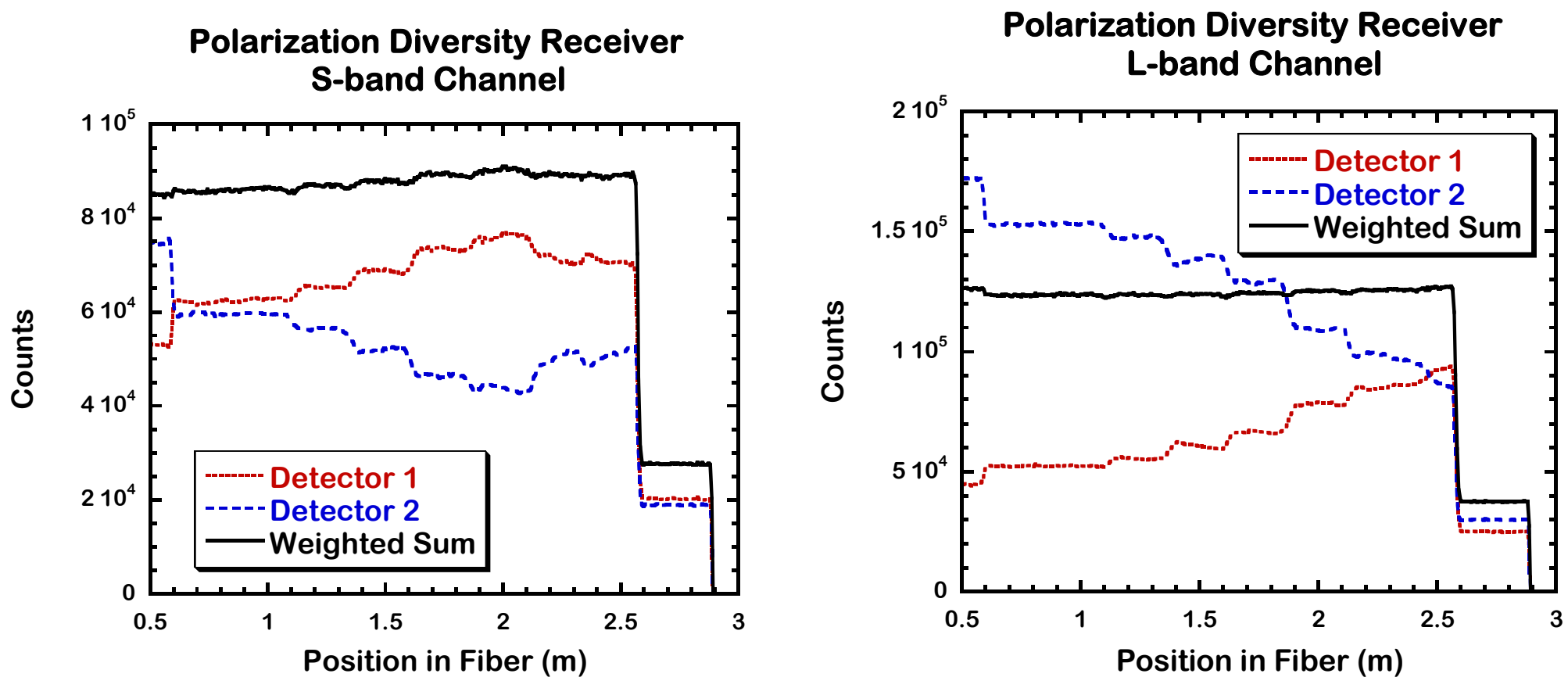

(a) (b)

Fig. 6. Plots of measured histogram data obtained with a polarization diversity receiver, showing the raw histogram data for the "H" and "V" polarizations, as well as the weighted sum of the two polarizations. Note that the weighted sum data are much more smooth (less fluctuations created by fiber bends) than the raw histograms for the two polarizations. (a) Histograms from the S-band channel (1460 – 1490 nm). (b) Histograms from the L-band channel (1570 – 1610 nm).

In the raw histogram data sets of Fig. 6, we see sharp changes in the total counts at the approximate locations of each of the five periodic bends in the fiber meander. From Fig. 2, we expect that the fiber meander bends would be periodic with a spacing of approximately 20 cm. There is an additional fiber bend at the extreme far end of the sensing fiber that is not shown in Fig. 2; this bend is created by the geometry required to immerse the sensing fiber endface into a container of index-matching fluid to suppress Fresnel backreflections from that endface.

To calculate the temperature of the fiber under test, we first calculated a weighted sum of the histograms from the two polarizations. The weighting factor is determined by the ratio of the detection efficiency (DE) of the two SNSPDs used to record the histograms for the "H" and "V" polarization states. The weighted-sum histograms are also shown in Fig. 6, and it is clear that the weighted-sum histograms are significantly less sensitive to the bend-induced polarization effects than the raw histograms from of the two polarizations.

From the weighted histogram data for the S-band and L-band channels, we then calculate the temperature of the fiber by use of Eq. (5) above. The resultant temperature as a function of position in the fiber is shown in Fig. 7. Also in Fig. 7 is a plot of the fiber temperature versus position when the polarization diversity receiver is not used (measurement system shown in Fig. 1 was used instead). In this figure we have indicated the approximate locations of the fiber meander bends. The data were taken at room temperature, so we expect a flat temperature profile. The PDR has significantly reduced the undesired fluctuations in the temperature versus position in the fiber. However, the temperature result of the PDR example still exhibits unwanted fluctuations. The higher-frequency fluctuations are likely noise-induced, as described in the uncertainty analysis, and could be reduced with higher pump power or longer integration periods. The lower-frequency fluctuations are more likely the result of residual polarization effects. We hypothesize that these fluctuations might result from the fact that we did not have four SNSPDs available for this experiment, so there was a short delay (a few minutes) and a change of fiber connections between the measurements of S-band and L-band channels, which might result in some residual polarization fluctuations.

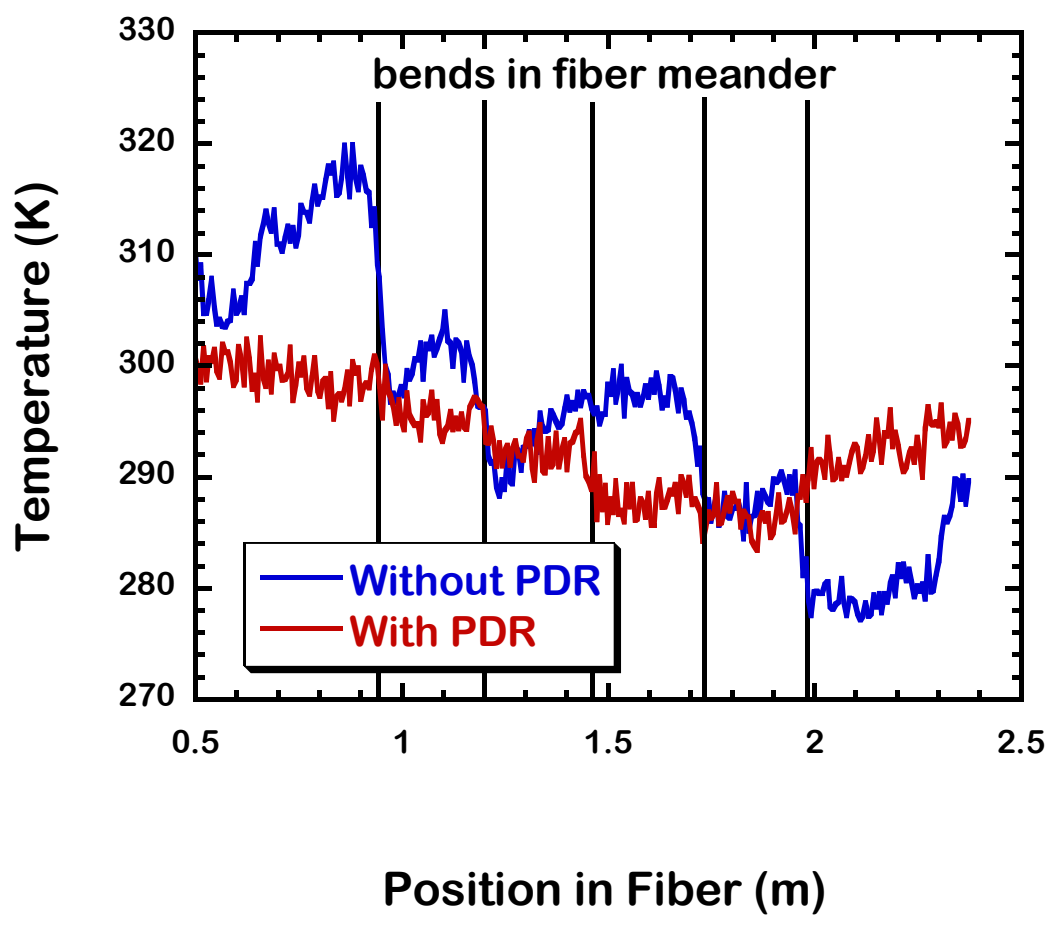

Fig. 7. Plot of measured temperature versus position in the sensing fiber. The sample was the two-dimensional fiber meander described above. For this particular measurement, the entire system was held at room temperature. Measurement was performed both with and without a polarization diversity receiver (PDR).

## 7. Conclusions

We have demonstrated a distributed fiber-optic temperature sensor with extremely high spatial resolution, as a result of the unique properties of our SNSPDs. The low timing jitter of the SNSPDs (on the order of 65 ps) enables spatial resolution on the order of 1 cm. Early demonstrations of distributed temperature sensors used multimode fiber to increase the collection efficiency of the Raman scattered photons, but here we are able to demonstrate a sensing system with standard telecom single-mode optical fiber. We expect that our technique could be applied to extremely long-range distributed sensors, because it is compatible with low-loss single-mode fiber and therefore unaffected by multimode fiber dispersion. Previously, we detailed the modifications necessary to achieve range on the order of 1 km while maintaining the low temperature uncertainty and high spatial resolution [7].

Our measurement system is capable of achieving extremely low temperature uncertainties: on the order of 3 K for a measurement integration period of 1 minute. The temperature uncertainty is currently limited by statistical fluctuations and it can be improved by increasing the integration period. Improvements in the temperature uncertainty are also possible with higher pump power, lower loss filters and/or higher efficiency SNSPDs. Alternatively, the temperature uncertainty can be improved by performing spatial averaging of existing data, sacrificing spatial resolution for reduced uncertainty in temperature. For example, with the data in Fig. 4, temperature uncertainty of the order of 1 K can be achieved with a spatial resolution of 5 cm.

We have shown that this measurement is also affected by systematic uncertainties, such as the polarization variations in the backscattered light. We demonstrated improved measurement performance with a polarization diversity receiver. This measurement could potentially be applied as a primary thermometer, but a full characterization of the properties of the laser, filters, detectors, as well as a detailed characterization of all of the systematic uncertainties would be necessary.

**Acknowledgements:** The authors are grateful to the National Institute of Information and Communications Technology (NICT) of Japan for providing SNSPDs for this experiment. MGT and RHH acknowledge funding from the UK Engineering and Physical Sciences Research Council. RHH also acknowledges a Royal Society University Research Fellowship. The authors thank Robert Maier and Bill MacPherson for useful discussions.